\documentclass[aps,pra,reprint,showpacs,longbibliography,superscriptaddress,groupedaddress]{revtex4-1}
\usepackage{xcolor}
\usepackage{amsmath}
\usepackage{hyperref}
\usepackage{graphicx}
\usepackage{braket}
\usepackage{textcomp}
\usepackage{pgfplots}
\usepackage{float}
\usepackage{soul} 
\usepackage{verbatim} 	
\usepackage[per-mode=symbol, product-units=single]{siunitx}
\DeclareSIUnit\bar{bar}
\DeclareSIUnit\G{G}
\newcommand*\circled[1]{\tikz[baseline=(char.base)]{
            \node[shape=circle,draw,inner sep=1.2pt] (char) {#1};}}

\begin{document}

\title{Compact detector for atom-atom correlations on an atom chip}
\author{Conny Glaser}\email{conny.glaser@uni-tuebingen.de}
\author{Dominik Jakab}
\author{Florian Jessen}
\author{Manuel Kaiser}
\author{J{\'o}zsef Fort\'agh}
\author{Andreas G\"unther}\email{a.guenther@uni-tuebingen.de}
\affiliation{Center for Quantum Science, Physikalisches Institut, Eberhard Karls Universit\"at
	T\"ubingen, Auf der Morgenstelle 14, D-72076 T\"ubingen, Germany}

\date{\today}

\begin{abstract}
We present a compact, ionization-based detector for the state-selective and spatially resolved measurement of individual Rydberg atoms trapped in the vicinity of an atom chip. The system combines an electrostatic lens system for guiding charged particles with an array of channel electron multipliers (CEMs) capable of detecting both ions and electrons produced by ionization. Designed for quantum information applications, this device enables the detection of correlations between spatially separated Rydberg qubits. Additionally, the electrodes provide compensation for stray electric fields and control over particle trajectories. The imaging system achieves a total magnification of more than 12, with a single-axis magnification up to 200, while maintaining low aberrations. We characterize the performance of the system using a charged particle trajectory simulation software and discuss how a coincidence measurement of ions and electrons can be used to calibrate the detection efficiency. This detector enables high-fidelity measurement of multiple Rydberg atoms and is well-suited for applications in cavity-mediated quantum gates.
\end{abstract}

\maketitle
\section{Introduction}\label{sec:introduction}
Neutral atoms excited to high-lying Rydberg states \cite{gallagher1994rydberg} have emerged as a powerful platform for quantum information processing. They offer strong and tunable interactions that enable high-fidelity quantum gates and scalable architectures \cite{lukin2001dipole, saffman2010quantum}. Recent advances in optical tweezer arrays and Rydberg-mediated entanglement have demonstrated fast two-qubit gates \cite{kaufman2021quantum, bluvstein2022quantum, graham2022multi}, coherent many-body dynamics \cite{scholl2021quantum, ebadi2021quantum, chen2023continuous}, and programmable quantum simulations of spin models \cite{schauss2015crystallization, labuhn2016tunable, bernien2017probing}. The ability to detect individual Rydberg atoms with high spatial and state selectivity is essential not only for verifying gate operations, but also for probing quantum correlations, calibrating interaction strengths, and studying hybrid quantum systems involving, for example, microwave cavities \cite{stammeier2017measuring, morgan2020coupling, kaiser2022cavity, kondo2024multiphoton}.

While optical detection techniques such as fluorescence imaging and Rydberg-EIT are routinely employed in neutral atom experiments \cite{pritchard2010cooperative, pritchard2013nonlinear, karlewski2015state}, these methods are often limited by detection sensitivity, speed, or spatial resolution. State-selective field ionization (SSFI), in contrast, provides a robust and high-fidelity approach to distinguish Rydberg levels by mapping atomic states to ionization signals \cite{gallagher1977field, van2006simultaneous, gregoric2018improving}. However, most existing SSFI implementations are designed for large ensembles or single-particle detection and lack the spatial discrimination necessary to resolve multiple Rydberg atoms simultaneously in a quantum computing context.

Our system integrates a superconducting atom chip for magnetic trapping of ground-state atoms \cite{fortagh2007magnetic}, controlled Rydberg excitation, and a set of electrodes for state-selective field ionization and guiding of the resulting charged particles through an electrostatic imaging system. These particles are detected at the image plane by an array of channel electron multipliers (CEMs) enabling the simultaneous detection of spatially separated Rydberg atoms with sufficient temporal resolution. The detector is specifically tailored for applications in quantum computing, where pairs of Rydberg atoms interact via a microwave cavity and require correlated, state-resolved measurements. Segmentation of the electrodes provides additional flexibility, enabling compensation of stray electric fields at the position of the Rydberg atoms and control over the particle trajectories towards the detectors. This approach enables precise, parallel readout of multi-atom Rydberg states, opening new possibilities for studying cavity-mediated interactions, quantum entanglement, and hybrid atom–photon platforms.

The paper is organized as follows. In Sec.~\ref{Sec_IonOptics}, we describe the design of the charged particle optics, including the layout of the electrodes, the materials considered for the detector assembly, the simulation method, and the theoretical basis for segmenting the electrodes. We also demonstrate how stray fields at the location of the atoms can be compensated. Sec.~\ref{sec:characterization} presents a detailed characterization of the system based on charged particle trajectory simulations. We analyze the electrostatic imaging system using a multipole expansion of the applied potentials. Starting with the monopole term, we examine the system’s magnification; the dipole term is used to laterally shift the extraction region and the resulting particle trajectories at the imaging plane. The quadrupole term is employed to enhance magnification along a single axis and to introduce an effective rotation of the object in the image. In Sec.~\ref{sec:aberrations}, we discuss image aberrations and examine the influence of magnetic fields on the system performance. Finally, in Sec.~\ref{sec:elion}, we describe a calibration procedure for measuring the detection efficiency of the system.

\section{Charged particle optics}\label{Sec_IonOptics}

\subsection{Design}
\begin{figure}
\centerline{\scalebox{1}{\includegraphics[width=0.5\textwidth]{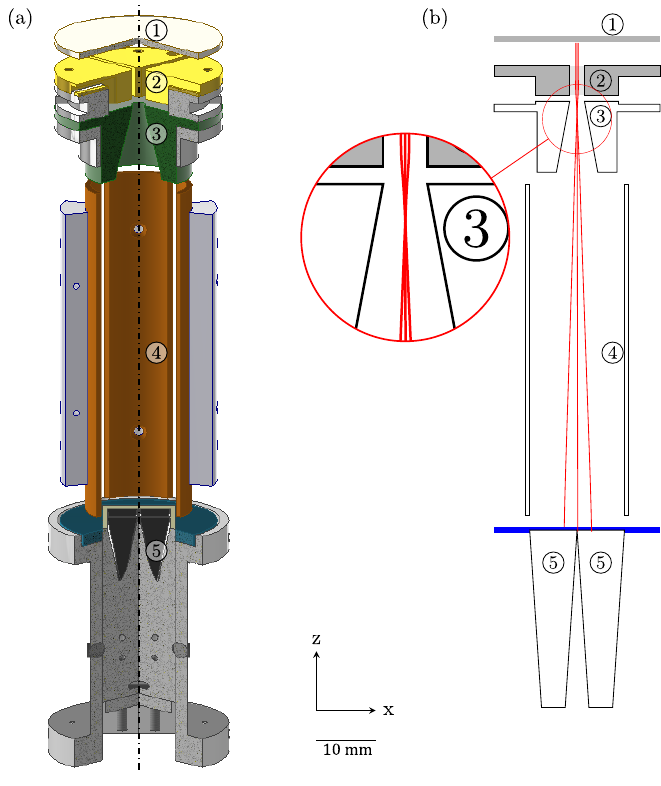}}}
	\caption{\textbf{Setup and Design.} (a) Technical setup of the ion-optical imaging system, consisting of several voltage supplied electrodes - from top to bottom: \protect\circled{1} atom chip, \protect\circled{2} extraction electrode, \protect\circled{3} conical electrode, \protect\circled{4} deflection electrodes (drift tube) and \protect\circled{5} CEM-array. All electrodes are isolated from each other and are mounted in a grounded tube (not shown). The black dashed-dotted line shows the position of the ion-optical axis, which is aligned perpendicular to the chip surface along the symmetry axis of the electrodes. (b) Trajectories for three ions starting at a mutual separation of \qty{100}{\um} and a distance of \qty{100}{\um} below the chip surface. Strong field gradients between the extractor and the conical electrode make for a lens effect (see inset). After being focused, the trajectories diverge towards the detection plane. Here, the voltages at the chip, the extractor, the conical electrode, the drift tube and the CEM entrance have been set to $U_{\text{chip}}= \qty{0}{\volt}$, $U_{\text{ext}}= \qty{-50}{\volt}$, $U_{\text{con}}= \qty{-1000}{\volt}$, $U_{\text{dt}}= \qty{-2000}{\volt}$ and $U_{\text{CEM}} =\qty{-2300}{\volt}$, respectively.}
\label{IonOptics}
\end{figure}

The trajectories of charged particles can be manipulated by means of electromagnetic fields \cite{drummond1984ion}. They can be used to realize electrostatic lenses and be combined to an ion-optical imaging system. Our setup is shown in Fig.~\ref{IonOptics}. It consists of several electrodes and aims to detect charged particles starting close to an atom chip surface. With the proposed detection system being placed in an ultrahigh vacuum (UHV) chamber, it is possible to detect both electrons or ions from the ionization process. As charged particles take the exact same trajectories in a static electric field regardless of the mass \cite{drummond1984ion}, the discussion is limited to the detection of rubidium ($^{87}$Rb) ions (unless noted otherwise). Starting close to the atom chip surface, they first pass a segmented extraction electrode and a conical electrode, which form an electrostatic lens \cite{sise2007aberration, el1971analysis, adams1972II}, typically used for low-energy ions \cite{drummond1984ion}. Subsequently, they propagate through a drift tube with deflecting electrodes and are detected by a set of four CEMs, arranged in a 2x2-array. 
The whole system extending from the chip surface up to the front facet of the CEM array features a compact size with \qty{28}{\mm} diameter and a total length of \qty{83}{\mm}.

The \textbf{atom chip} \circled{1} (voltage $U_{\text{chip}}$) sets the reference potential for the imaging system. It may host several lithographically implemented structures to manipulate the atoms \cite{fortagh2007magnetic}. Applying a voltage may either be achieved by design or by covering the chip surface with a high resistive conducting layer. Most conveniently, the reference potential is set to ground $U_{\text{chip}}= \qty{0}{\volt}$.

The \textbf{extraction electrode} \circled{2} (voltage $U_{\text{ext}}$), of \qty{1}{\mm} thickness and a central bore of \qty{2.5}{\mm} diameter, is placed \qty{4.5}{\mm} below the chip surface. Applying a voltage to the extraction electrode, a homogeneous $z-$directional field at the position of the atoms is generated. The voltage is tunable and ramps the electric field to ionize the Rydberg atoms. The ion is then accelerated towards the extraction electrode, passing its central bore. The extraction electrode is segmented into four quadrants (cf.~Fig.~\ref{IonOptics}a) allowing compensation for detrimental field components, such as electric fields caused by adsorbates sticking to the chip surface \cite{mcguirk2004alkali, Obrecht2007, tauschinsky2010spatially, hattermann2012detrimental, chan2014adsorbate}. The electric dipole and quadrupole field terms originating from the segmentation can be further used to increase the detection area and the functionality of the imaging system, as discussed in Sec.~\ref{sec:characterization}. 

With the \textbf{conical electrode} \circled{3} (voltage $U_{\text{con}}$) entrance placed 1\,mm away from the extractor electrode and with largely different potentials at both electrodes, a region of strong field gradients is generated. It acts as an electrostatic lens for ions. As depicted in the inset of Fig.~\ref{IonOptics}b, the initially parallel ion trajectories are focused in the lens' focal plane and diverge afterwards. As the name suggests, this electrode is conically shaped, allowing all particles from the ionization region to pass through.

The \textbf{drift tube} \circled{4} (voltage $U_{\text{dt}}$) of \qty{56}{\mm} length is set to a potential close to the CEMs entrance voltage and is used to increase the imaging magnification by allowing the particle trajectories to further expand after leaving the conical electrode. It is also segmented into four pieces, called deflector electrodes, to allow for positioning the ion trajectories on the subsequent CEMs and to introduce extremely high single-axis magnifications (cf.~Sec.~\ref{subsec:SingleAxis}).

A set of four \textbf{CEM}s \circled{5} (voltage $U_{\text{CEM}}$) with opening apertures of \qtyproduct{5x5}{\mm} and \qtyproduct{5x10}{\mm} complete the ion-optical setup. They are arranged in a 2x2 array with a total size of \qtyproduct{10x15}{\mm}. To reduce thermal load on the system and to optimally use the detection area, we have chosen high resistive CEMs with a wall thickness of \qty{200}{\um} \cite{Sjuts}, leaving gaps of only \qty{400}{\um} between the individual CEMs. Optimally, the CEM front facets are set to a voltage close to the drift tube voltage. 
Ions hitting the resistive surface of the CEM produce secondary electrons, which are then accelerated and further amplified towards the CEM output, typically set to \qty{-100}{\volt} for ion detection. With a potential difference of \qty{100}{\volt} \cite{Sjuts}, the electron pulse is guided to the grounded detection anode, where it produces a \qty{20}{\ns} long charge pulse of about $10^7$ electrons. The charge pulse is electronically amplified and detected with a digital counter module, by which the incoming particles can be detected with a temporal precision better than \qty{100}{\ps} \cite{SwabianInstruments}. To ensure a smooth potential landscape around the CEMs entrance, the CEM array is mounted into a circularly shaped electrode (shown in blue in Fig.~\ref{IonOptics}), which is set to the CEMs front potential. If additionally, the potential difference between the deflection electrodes and the CEM entrance is kept small, $\left|U_{\text{dt}}-U_{\text{CEM}}\right|\ll \left|U_{\text{CEM}}\right|$, the potential gradient towards the CEM output will be much stronger than back to the deflection electrodes. This ensures that secondary electrons will most likely be guided to the CEM output and not back to the ion-optical system, otherwise the detection efficiency near the CEM edges is strongly reduced. The detection efficiency in general is determined by the kinetic energy of the incoming particle, which in our case is directly set by $U_{\text{CEM}}$, and the voltage drop across the CEM. The latter is typically chosen above \qty{2}{\kilo\volt}, to ensure a sufficient gain of the electron avalanche. This leaves the electron emission probability on the initial particle impact as main limitation for the quantum efficiency of the CEM \cite{henkel2011photoionisation}.
To achieve efficiencies $> 50\%$, our detectors require at least \qty{2}{keV} kinetic energy for ions and \qty{20}{eV} for electrons \cite{Sjuts}. While individual detection of two spatially separated particles would require a minimum of two CEMs, the system with four CEMs offer directional flexibility. Fine tuning of the individual ion trajectories onto the CEMs is enabled via the segmented deflection electrodes of the drift tube. In principle, the CEM array can be replaced by a multi-channel plate (MCP), which allows for spatially localized production of secondary electrons. Using two MCPs in chevron configuration, similar pulse amplitudes as for the CEMs are achieved \cite{wiza1979microchannel}. In this case an additional detection anode, capable of detecting the spatially localized electron pulse is required. Possible solutions are ranging from quadrant anodes \cite{lampton1976quadrant}, over cross-strip anodes \cite{siegmund2001cross} up to delay-line-anodes \cite{keller1987position}. In special configurations these systems also allow for simultaneous detection of multiple particles \cite{jagutzki2002multiple}. However, these MCP based systems typically lose spatial resolution for particles with small temporal separation, which is most crucial for detecting two particles at the same time. Therefore we have chosen separate detectors (CEMs) for the individual particles.

\subsection{Materials \& Constraints}\label{subsec:constraints}
Isolating the different electrodes from each other and from other conducting elements in the UHV chamber is of specific importance to prevent flashovers and short circuits. In principle, all conducting parts can be machined from oxygen free high conductivity (OFHC) copper or stainless steel and all insulating parts from thermoplast polyether ether ketone (PEEK) or a machinable glass-ceramic such as MACOR, which are all vacuum compatible. However, due to its temperature and outgassing performance, PEEK is only used in cryogenic environments, where good vacuum conditions can be achieved after moderate baking at temperatures below \qty{140}{\celsius}. At higher temperatures PEEK undergoes a glass transition \cite{cheng1986glass} and is not suitable for vacuum applications, such that MACOR would be used instead. Both, PEEK and MACOR, feature a dielectric strength above \qty{20}{\kV\per\mm} \cite{drummond1984ion}, which is negligible compared to the vacuum breakdown voltage and insulator surface effects. In an ultrahigh vacuum chamber at a pressure below \qty{1e-10}{\milli\bar}, the vacuum breakdown voltage amounts \qty{6}{\kV\per\mm} \cite{drummond1984ion}. Taking leakage currents via the insulator's surface into account, a surface tracking distance of \qty{1}{\mm\per\kV} is required \cite{drummond1984ion}. In order to avoid voltage breakdowns the system is designed to have minimal separation of \qty{1}{\mm} between electrodes and maximal voltage differences of \qty{1}{\kilo\V\per\mm}. All ion-optical electric parts are also electropolished to reduce surface roughness and increase breakdown voltages. Additionally, the detector system is encapsulated in a grounded tube, to allow for simple handling and installation of the imaging system to different experimental apparatuses. Moreover, surface charges at the isolating surfaces have to be avoided, as they might perturb the electrostatic fields and thus the ion trajectories. Therefore, all electrodes are designed in a way, that no point along the ion trajectory has a direct line of sight to an insulating surface.

\subsection{Ion-optics Simulation}
To calculate the field and particle trajectories a commercial charged particle optics simulation program (SIMION) is used in combination with MATLAB and Python scripts. For a given electrode geometry SIMION calculates the electrostatic potential and field by solving the Laplace equation with a finite difference method. Charged particle trajectories are then calculated by numerically solving the Newtonian equation of motion in this field. The correlation detector is designed to resolve two spatially separated Rydberg atoms close to an atom chip surface. Although, the exact distance between atoms and chip surface is largely irrelevant for the simulations, we have chosen an experimentally realistic value of \qty{100}{\um} for all simulations. Figure~\ref{IonOptics}b shows the exemplary trajectories for three particles starting \qty{100}{\um} to the chip surface with separations of 0 and \qty{\pm100}{\um} to the ion-optical axis (black dashed line in Fig.~\ref{IonOptics}a) for typical ion detection voltages of $U_{\text{chip}}=\qty{0}{\V}$, $U_{\text{ext}}=\qty{-50}{\V}$, $U_{\text{con}}=\qty{-1000}{\V}$, $U_{\text{dt}}=\qty{-2000}{\V}$ and $U_{\text{CEM}}=\qty{-2300}{\V}$.
In this configuration the ions require about \qty{2.45}{\micro\second} to reach the detector. At same, but inverted, voltage settings, electrons would be faster by a factor of $\sqrt{m_{\text{Rb}}/m_{e}} \approx 400$ with the corresponding ion and electron mass $m_{\text{Rb}}$ and $m_{e}$, respectively. However, for electron detection typical voltages are lower, $U_{\text{chip}}=\qty{0}{\V}$, $U_{\text{ext}}=+\qty{50}{\V}$, $U_{\text{con}}=+\qty{1000}{\V}$, $U_{\text{dt}}=+\qty{100}{\V}$ and $U_{\text{CEM}}=+\qty{120}{\V}$, resulting in a transit time of a few \qty{}{\nano\second}. 

\subsection{Segmented Electrodes}
\begin{figure}[tbp]
	\includegraphics[width=0.47\textwidth]{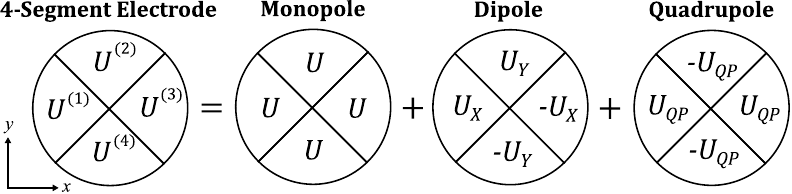}
	\caption{\textbf{Segmented Electrode.} The field and/or voltage on a segmented electrode can be decomposed to a superposition of three contributions: The first being the monopole term with identical voltage on all segments; the second being the dipole term with same voltage magnitude but different signs on opposing electrodes; and the last being the quadrupole term with identical voltage magnitude on all segments, but changing signs on neighbouring segments.}
	\label{fig:cem_segments}
\end{figure}
With both the extraction and deflection electrode being segmented into four quadrants, they allow for rich influence onto the electrostatic potential and the corresponding particle trajectories. We shall model this by a simplified electrode that has a circular profile and is symmetrically cut into four pieces, each supplied with an individual voltage $U^{(i)}_{\text{el}}$ with $i=1-4$ denoting the segment and the lower index optionally denoting the electrode. Most general, the field generated by such an electrode can then be decomposed into various multipoles, with the corresponding decomposition for a four-segment electrode including only the monopole, dipole and quadrupole terms. Figure~\ref{fig:cem_segments} shows this decomposition in terms of the corresponding voltages at the individual segments, with
\begin{eqnarray}
        U &=& \frac{1}{4}\left(U^{(1)} + U^{(2)} + U^{(3)} + U^{(4)}\right)\\
        U_X &=& \frac{1}{2}\left(U^{(1)}-U^{(3)}\right),\, U_Y = \frac{1}{2}\left(U^{(2)}-U^{(4)}\right)\\
        U_{\text{QP}}&=&\frac{1}{4}\left[U^{(1)}+U^{(3)}-\left(U^{(2)}+U^{(4)}\right)\right].
\end{eqnarray}

The first contribution resembles the monopole term, with identical voltages $U$ on all segments. Thereby, $U$ is given as mean voltage across all segments and the resulting field close to the center of the electrode being homogeneous and having only a component perpendicular to the electrode ($z$-direction), but no component in parallel ($xy$-plane)
\begin{equation}
    \vec{E}\sim\left(\begin{array}{c}0\\ 0\\ U\end{array} \right).
\end{equation}
However, this changes when moving along the optical axis, where the strength of the z-field usually varies due to the finite geometry of the electrode. Following Gauss's theorem, radial symmetric fields perpendicular to the optical axis, $E_r$, must then emerge \cite{bergmann2004optik}:
\begin{equation}
    \vec{E}_r\sim -\frac{r}{2}\frac{d E_z(0,0,z)}{dz}\vec{e}_r,
    \label{eq:focusEffect}
\end{equation}
with $r=\sqrt{x^2+y^2}$ denoting the distance to the optical axis and $\vec{e}_r=(x,y,0)/r$ being a unit vector pointing in radial direction. Equation \eqref{eq:focusEffect} sets the base for the principal lensing effect in our ion-optical system.

The second contribution is the dipole term resulting from opposing segments having the same voltage magnitude but different signs (cf.~Fig.~\ref{fig:cem_segments}). Close to the center of the electrode this results in homogeneous fields in $x$- and $y$-direction 
\begin{equation}
    \vec{E}\sim \left(\begin{array}{c}U_X\\ U_Y\\ 0\\\end{array} \right)=  U_{XY}\left(\begin{array}{c}\cos\alpha\\ \sin\alpha\\ 0\\\end{array} \right),\label{eq:dipole1}
\end{equation}
which can be understood as single homogeneous field component of strength $E_{XY}\sim U_{XY}$ in a direction described by the angle $\alpha$ with the $x$-axis.
\begin{equation}
    U_{XY}=\sqrt{U_X^2+U_Y^2},\quad \alpha=\tan^{-1}\left(\frac{U_Y}{U_X}\right).\label{eq:dipole2}
\end{equation}
In our system this field is typically used to displace ion trajectories in distinct directions.  

The third and last contribution is the quadrupole term, with all segments having the same voltage magnitude $U_{\text{QP}}$ but changing sign between neighbouring segments (cf.~Fig.~\ref{fig:cem_segments}). The corresponding field close to the electrode center is that of a typical planar quadrupole
\begin{equation}
    \vec{E}\sim U_{\text{QP}}\left(\begin{array}{c}-x\\ y\\ 0\\\end{array} \right),
\end{equation}
i.e. inhomogeneous with increasing field strength from the center. Similar to the radial monopole term in Eq.~\eqref{eq:focusEffect}, this contribution can be used for focusing the beam in either the $x$- or $y$-direction, while defocusing it in the other direction.

In Sec.~\ref{sec:characterization}, the effects and functionalities of these terms on the ions' trajectories in the proposed detector system with the segmented extractor and deflector electrodes are discussed in more detail. However, the monopole and dipole terms of the extractor can also be used beyond ion trajectory manipulation to generate electric fields at the atomic position close to the chip surface.

\subsection{Stray field compensation}\label{subsec:strayFieldCompensation}

The electric field of the extractor can provide a means of compensating for lateral stray fields, such as detrimental fields from the atom chip surface. This is particularly important for Rydberg atoms, which are very sensitive to any residual field. In this context, the monopole term of the extractor can be used during the Rydberg excitation and Rydberg interaction phase (i.e. outside the ion-detection phase) to compensate for fields in the $z$-direction (perpendicular to the chip surface) with a field strength of \qty{2.2}{\volt\per\cm} per volt at the electrodes. Similarly, the use of dipole terms allows compensation of lateral fields parallel to the chip surface with a field value of \qty{26.5}{\milli\volt\per\cm} per volt difference at opposite electrodes. According to Eq.~\eqref{eq:dipole1} and \eqref{eq:dipole2} the angle $\alpha$ of the lateral field can be arbitrarily tuned.

\section{Characterization}\label{sec:characterization}
The focus of the characterization lies on the overall magnification of the detector system, as well as on the possibilities to influence the ion trajectories via the segmented extraction and deflection electrodes. All those properties are partially influenced by aberrations, which are discussed in Sec.~\ref{sec:aberrations}. The imaging quality in terms of resolution and sharpness are of secondary importance and shall not be discussed in detail. 

\subsection{Image Magnification}\label{subsec:magn}
The magnification of the system plays a critical role in determining the minimum spatial separation required for the simultaneous detection of particles. While the overall system magnification — addressed in detail below — provides a general threshold for resolvable distances, the single-axis magnification specifically characterizes the minimum resolvable separation in the case of two particles. This case is examined in detail in Sec.~\ref{subsec:SingleAxis}.

\begin{figure}[tbp]
	\centerline{\scalebox{1}{\includegraphics[width=0.48\textwidth]{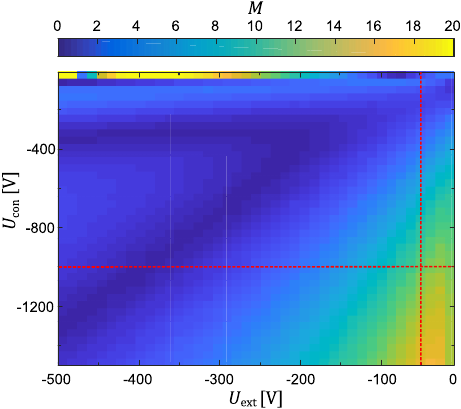}}}
	\caption{\textbf{Magnification.} Image magnification close to the optical axis as function of the voltages on the extractor $U_{\text{ext}}$ and the conical electrode $U_{\text{con}}$ for particles starting on the optical axis at a distance of 100\,\textmu m below the chip surface. The remaining voltages have been fixed to $U_{\text{chip}}=\qty{0}{\V}$, $U_{\text{dt}}=\qty{-2000}{\V}$ and $U_{\text{CEM}}=\qty{-2300}{\V}$, respectively. The intersection of the red dashed lines denote the working point used in this work with $U_{\text{ext}}=\qty{-50}{\V}$, $U_{\text{con}}=\qty{-1000}{\V}$ and $M=\num{12.25}$.}
	\label{fig:M_ext_con}
\end{figure}

The monopole terms of the ion optical electrodes define the overall magnification of the detector system. The major part of the magnification occurs at the interspace between the extractor and the conical electrode in conjunction with the length of the imaging system. Different voltages result in different magnifications at the imaging plane (front facet of the CEM array), as shown in Fig.~\ref{fig:M_ext_con}. 
In this case, the magnification is determined by placing the ions in a square lattice symmetrically around the optical axis, with a lattice constant and distance to the chip surface of \qty{100}{\um}. The voltages of the chip, the deflector electrodes and the CEM entrance are fixed to $U_{\text{chip}}=\qty{0}{\V}$, $U_{\text{dt}}=\qty{-2000}{\V}$ and $U_{\text{CEM}}=\qty{-2300}{\V}$, respectively. The magnification is then given by the ratio of the particle separation at the imaging plane to the initial separation.
Figure~\ref{fig:M_ext_con} shows that large field gradients are favorable for achieving high magnifications. These can be realized by either high extractor and low conical voltages (upmost region in Fig.~\ref{fig:M_ext_con}) or low extractor and high conical voltages (lower right region in Fig.~\ref{fig:M_ext_con}). The latter configurations are preferable as they are less sensitive to voltage variations and fluctuations, as can be seen from the gradients of the magnification in the aforementioned areas of the plot. With this configuration and for the voltages given in Fig.~\ref{fig:M_ext_con} the maximal magnification amounts to 16. To avoid effects due to stray electric fields, the acceleration voltage at the extractor is limited to a minimum magnitude of 50\,V. As constrained by breakdown and design limits (see Sec.~\ref{subsec:constraints}), the maximal voltage difference between the extractor and conical electrodes is limited to \qty{1000}{\V}. This yields a maximal magnification of about 12, as realized for $U_{\text{ext}}=\qty{-50}{\V}$ and $U_{\text{con}}=\qty{-1000}{\V}$.
 
\begin{figure}[tbp]
\includegraphics[width=0.47\textwidth]{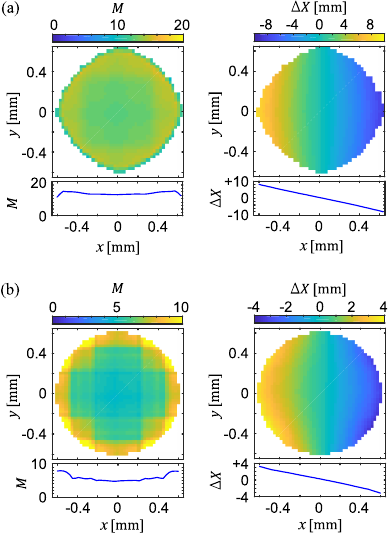}
\caption{\textbf{Image properties.} Magnification $M$ and image displacement $\Delta X$ across the imaging plane for (a) $U_{\text{ext}}=\qty{-50}{\V}$  and (b) $U_{\text{ext}}=\qty{-200}{\V}$. $M$ and $\Delta X$ are calculated for ions starting at different displacements $x$ and $y$ to the optical axis with a distance of \qty{100}{\um} to the chip surface. The remaining voltages have been fixed to $U_{\text{chip}}=\qty{0}{\V}$, $U_{\text{con}}=\qty{-1000}{\V}$, $U_{\text{dt}}=\qty{-2000}{\V}$ and $U_{\text{CEM}}=\qty{-2300}{\V}$. For symmetry reasons, the image displacement $\Delta Y$ (not shown) is identical to $\Delta X$ for exchanged starting positions $x$ and $y$.}
\label{fig:FieldOfView}
\end{figure}

For particles starting displaced from the optical axis, the magnification is slightly increased towards the edges, as shown in Fig.~\ref{fig:FieldOfView} for extractor voltages of \qty{-50}{\V} and \qty{-200}{\V}, respectively.
At the lower voltage of $U_{\text{ext}}=\qty{-50}{\V}$ the central magnification amounts $M=12.25$ and increases up to $M=14.5$ at the edge of the detection plane (see Fig.~\ref{fig:FieldOfView}a). With an increasing voltage at the extractor the central magnification is reduced and the detection area is limited by the hole in the extraction electrode with $2.5$\,mm diameter. The magnification then drops to $M=5$ for $U_{\text{ext}}=\qty{-200}{\V}$ (see Fig.~\ref{fig:FieldOfView}b). Aside from the magnifications, Fig.~\ref{fig:FieldOfView} shows the position of the ions at the detector for particles starting displaced from the optical axis. For symmetry reasons only $\Delta X$ - the displacement along $x$ - is shown. Especially in the high magnification case (Fig.~\ref{fig:FieldOfView}a), which the ion-optics has been optimized for, the image displacement varies linearly across the detection plane with minimal crosstalk between the $x$- and $y$-direction. This results in negligible image skew across the entire detection plane.

\subsection{Detection and Extraction Region}\label{subsec:detection_extraction_region}
\begin{figure}[tbp]
\centerline{\scalebox{1}{\includegraphics[width=0.48\textwidth]{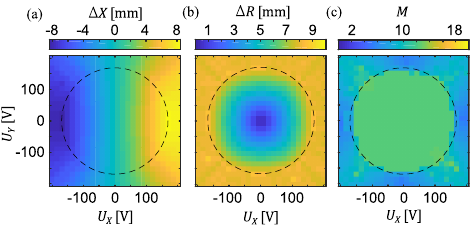}}}
	\caption{\textbf{Image deflection.} Ion impact positions for different offset voltages $U_X$ and $U_Y$ at the deflector electrodes: (a) displacement in $x$-direction and (b) displacement in radial direction. (c) Magnification of the imaging system as function of the offset voltages $U_X$ and $U_Y$ at the deflection electrodes. The magnification $M$ is calculated as mean of the corresponding $M_x$ and $M_y$ values, which have been extracted from two ions starting symmetrically around the optical axis with a separation of \qty{100}{\um} in $x$- and $y$-direction, respectively. For the simulations, the ion starting positions have been fixed to the optical axis at a distance of \qty{100}{\um} below the chip surface. The voltages of the remaining electrodes have been fixed to $U_{\text{chip}}=\qty{0}{\V}$, $U_{\text{ext}}=\qty{-50}{\V}$, $U_{\text{con}}=\qty{-1000}{\V}$, $U_{\text{dt}}=\qty{-2000}{\V}$ and $U_{\text{CEM}}=\qty{-2300}{\V}$. All particles within the dashed circle reach the detection plane.}
	\label{fig:Ablenker2}
\end{figure}

Next, the effects of the dipole term on both the deflector and the extractor electrodes are discussed. The impact position of the ions at the CEM-array is adjusted by deflecting the beam paths in lateral direction to the optical axis. This is realized by the deflection electrodes (cf.~\circled{4} in Fig.~\ref{IonOptics}). During normal operation the deflector segments are set to $U_{\text{dt}}\approx \qty{-2000}{\V}$, which guides the ions onto the CEM array. For beam deflection, dipole offset voltages $U_X$ and $U_Y$ are superimposed onto $U_{\text{dt}}$, with alternating signs for opposite electrodes as shown in Fig.~\ref{fig:cem_segments}. This allows for the deflection of the beam in the $xy$-plane, while keeping the imaging properties of the ion-optical system mainly unaffected. Figure~\ref{fig:Ablenker2} shows the expected ion impact positions, as simulated for ions starting on the optical axis at a distance of \qty{100}{\um} below the chip surface.
The voltages of the extractor and the conical electrode have been fixed to \qty{-50}{V} and \qty{-1000}{V}, respectively. The $\Delta X$-displacement in Fig.~\ref{fig:Ablenker2}a, shows a linear response to the applied voltages and can be addressed almost independently from the $\Delta Y$-displacement. Only at the very edge of the imaging plane, the linearity shows small, but negligible, deviations from its mean value of \qty{48}{\um\per\V}. Therefore, the radial displacement in Fig.~\ref{fig:Ablenker2}b is highly symmetric and shows a circular profile around the center of the imaging plane.
The same holds for the magnification, which is almost perfectly even across the whole range of deflection voltages. As seen in Fig.~\ref{fig:Ablenker2}c, it shows only negligible deviations from its mean value of 12.25 with deviations below $6.5\%$. The circularly shaped increase of magnification (brighter dots) at the very edges is due to the finite size of the CEM detection plane and therefore purely a geometrical limitation. Altogether, this makes the segmented deflection electrode a suitable tool for adjusting the ion trajectories onto the CEM-array without loss in imaging quality and high linearity in the image displacement.\\
\begin{figure}[tbp]
\centerline{\scalebox{1}{\includegraphics[width=0.48\textwidth]{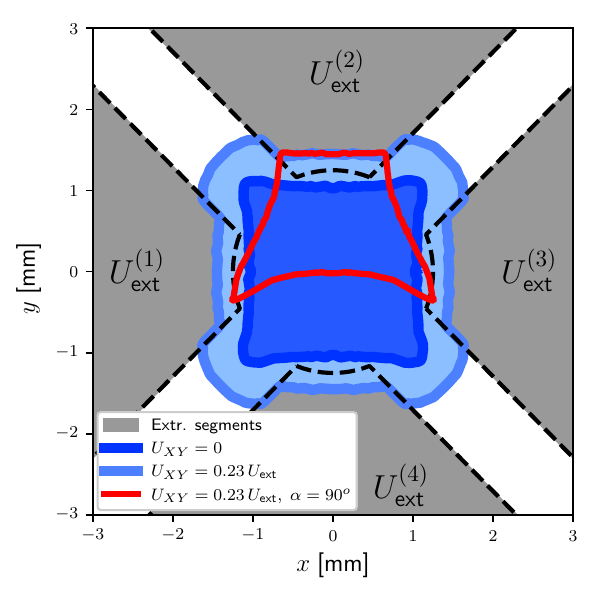}}}
\caption{\textbf{Extraction region.} Simulated extraction region of the detector. The black dotted border and grey areas show the boundaries of the segmented extraction electrodes. The normal extraction region, without dipole term applied to the extractor, is shown in dark blue. The increased region with additional dipole term, as averaged for all directions $\alpha$, is shown in light blue, with a specific realization for $\alpha=\ang{90}$ outlined in red. The plot was simulated using $U_{XY} = 0.23\,U_{\text{ext}}$ and applying voltages $U^{(1),(2),(3),(4)}_{\text{ext}}$ as calculated from Eq.~\eqref{eq:dipole1} to the extractor. The remaining voltages have been set to $U_{\text{chip}}=\qty{0}{\V}$, $U_{\text{ext}}=\qty{-50}{\V}$, $U_{\text{con}}=\qty{-1000}{\V}$, $U_{\text{dt}}=\qty{-2000}{\V}$ and $U_{\text{CEM}}=\qty{-2300}{\V}$.}
\label{fig:detectionArea}
\end{figure}

When dipole fields are used on the extraction electrodes instead, their applications are twofold: Firstly, they can provide compensation for lateral stray fields, as discussed in Sec.~\ref{subsec:strayFieldCompensation}. Secondly, dipole fields can be used to shift the extraction region to significantly increase the maximum detectable initial particle displacement from the optical axis. Thereby, the extraction region is defined as simply connected region in the object plane (\qty{100}{\um} from the chip surface) around the ion optical axis, for which the ion trajectories reach the detector. As the trajectories are highly tunable by the deflector electrodes, particles entering the deflector region (12.5mm into the detector) are considered to be detectable. In general, the extraction region is limited by the size of the hole in the extraction electrode, which in our case amounts to a diameter of $\qty{2.5}{\mm}$. Based on the simulations, this limits the effective extraction region to an almost square area of width $w_{0}\approx \qty{2.12}{\mm}$, as seen in Fig.~\ref{fig:detectionArea}. However, there are \qty{1}{\mm} gaps between the segments of the extractor electrodes. In these gaps the width is increased slightly, as there the outer trajectories that would normally hit the electrodes can pass between them. The extraction region can be shifted by taking advantage of the segmented extraction electrodes. For a particle starting displaced by coordinates ($x,y$) from the optical axis, outside the extraction region of width $w_0$, this requires that extractor potentials are modified by the dipole voltages $U_{X} = -U_{XY} \cos(\alpha)$ and $U_{Y} = -U_{XY} \sin(\alpha)$, with $\alpha = \tan^{-1}(y/x)$ and $U_{XY}$ being the base dipole voltage. $U_{XY}$ is set in the simulations so that the maximal detectable displacement is achieved while still being able to detect particles from the center of the extraction region for any $\alpha$. Figure~\ref{fig:detectionArea} shows the resulting size of the maximal extraction region for charged particles with $U_{XY} = 0.23\,U_{\text{ext}} = \qty{-11.5}{\V}$, where we took the farthest detectable particle for each $\alpha \in \{0, 2\pi\}$. Using this method, we find the maximal detectable area being increased by almost a factor of 2, with the maximal extraction region width being increased to $1.4 w_0 \approx \qty{2.94}{\mm}$.
However, it is important to note, that while this technique allows to detect particles outside the "normal" extraction region, it does so only in a smaller, shifted region (cf.~Fig.~\ref{fig:detectionArea} - red line for $\alpha = 90^o$). In principal it is possible to further increase the extraction region by employing dynamical tuning schemes for the electrode voltages. However, this requires case-specific simulations outside the scope of this article.

\subsection{Single-axis Magnification}\label{subsec:SingleAxis}
\begin{figure}
\centerline{\scalebox{1}{\includegraphics[width=0.48\textwidth]{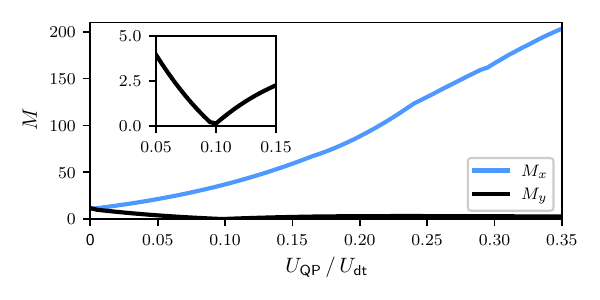}}}
\caption{\textbf{Single-axis magnification.} Simulation of the magnifications $M_x$ and $M_y$ in $x$- and $y$-direction, respectively, as derived from ion trajectory simulations for initial positions distributed within a circle of \qty{100}{\micro\meter} radius, around the optical axis. The $x$-axis shows the ratio between the base deflector voltage ($U_{\text{dt}} = \qty{-2000}{V}$) and the superimposed quadrupole voltage ($U_{\text{QP}}$), while $U_{\text{ext}} = \qty{-50}{V}$, $U_{\text{con}} = \qty{-1000}{V}$, $U_{\text{CEM}} = \qty{-2300}{V}$. The inset shows the local minimum of the $y$-axis magnification near $U_{\text{QP}}/U_{\text{dt}} = 0.1$, which can be interpreted as a focusing effect in the $y$-axis on the detection plane.}
\label{fig:HighXMagnification}
\end{figure}
In Sec.~\ref{subsec:magn}, the total magnification was shown to be tunable between $M=5-12.25$, mainly constrained by the applicable maximum voltage differences between the extractor and conical electrodes. This range can be further extended by applying a quadrupole lens effect \cite{hytch2022quadrupoles}, enabling an additional single-axis magnification. As a quadrupole field in the extraction region would change the detection area, the deflector electrodes are chosen for this purpose. Increasing the voltages on opposite deflector pairs by $U_{\text{QP}}$ while reducing them on the other pair allows strong anisotropic magnification of $M_{x/y}>200$, while keeping $M_{y/x}$ small. In theory, the achievable magnification is only constrained by the physical dimensions of the deflector region and imaging plane. In our detector system, applying a quadrupole term $U_{\text{QP}} = \qty{-700}{V}$ to the deflector base voltage $U_{\text{dt}}=\qty{-2000}{\volt}$, i.e. $x$-electrode voltages of $U^{(1),(3)}_{\text{dt}} = \qty{-2700}{\volt}$ and $y$-voltages of $U^{(2),(4)}_{\text{dt}} = \qty{-1300}{\volt}$, results in an $x$-axis magnification of $M_x \approx 203$, while $M_y < 3.2$. The relation between the applied voltage and the magnification is shown in Fig.~\ref{fig:HighXMagnification}. Considering the \qty{400}{\um} gap between the detectors, this single-axis magnification allows for the detection of particles separated by \qty{\sim 2}{\um} in our system. Apart from the high magnification, there is also a local minimum of the $y$-axis magnification near $U_{\text{QP}}/U_{\text{dt}}=0.1$, which can be used to implement high magnification on one axis, while focusing on the other.
\begin{figure}
\centerline{\scalebox{1}{\includegraphics[width=0.48\textwidth]{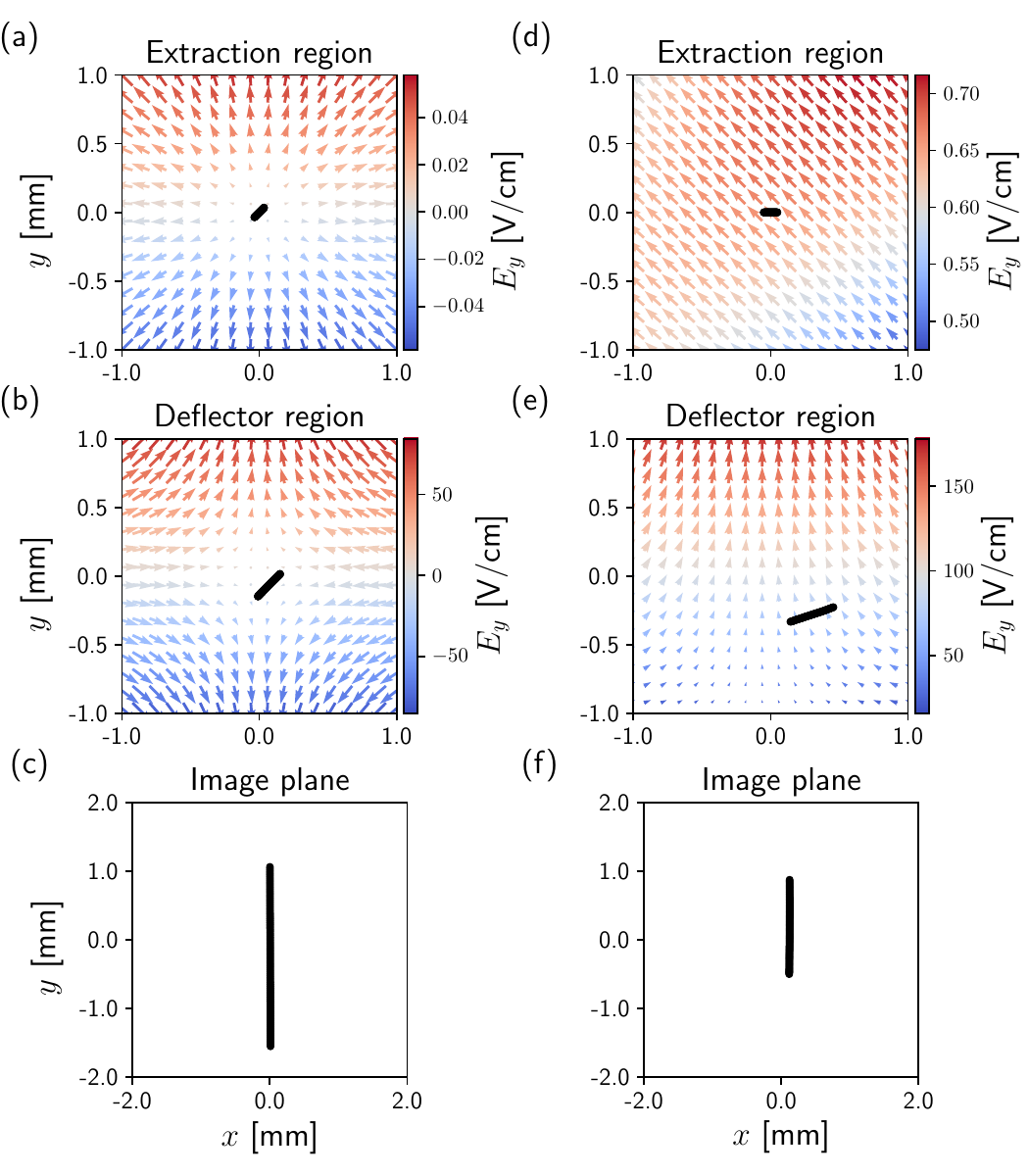}}}
\caption{\textbf{Image alignment.} (a) Atomic line distribution (black dots) with \qty{100}{\um} length and $\alpha=\ang{45}$ angle to the $x$-axis at the object plane (extractor region) in conjunction with the $xy$-component of the electric field (vectors) and the strength of the $y$-field component (color of vectors). The atom positions entering the deflector region are simulated in (b), with $E_{xy}$ shown inside the deflector, after applying a single-axis magnification with $U_{\text{QP}} = \qty{+200}{V}$ to the deflection electrodes. (c) Resulting image at the detectors. (d) In the extreme case of $\alpha=\ang{0}$, the extractor's dipole term can be used to introduce a spatial dependence of $E_{y}$ in the extraction region (see color change of vectors along x). (e) Resulting rotation at the entrance of the deflector region. (f) Using the deflector's dipole and quadrupole field, the image can be centered to the detector and aligned to the y-axis, respectively. The voltage settings for the simulations are: $U_{\text{chip}}=\qty{0}{\V}$, $U_{\text{ext}}=\qty{-50}{\V}$, $U_{\text{con}}=\qty{-1000}{\V}$, $U^{(1),(3)}_{\text{dt}} = \qty{-1800}{\V}$, $U^{(2),(4)}_{\text{dt}} = \qty{-2200}{\V}$ and $U_{\text{CEM}}=\qty{-2300}{\V}$ for \mbox{(a-c)} and $U^{(1),(2)}_{\text{ext}} = \qty{-55}{\V}$, $U^{(3),(4)}_{\text{ext}} = \qty{-45}{\V}$, $U^{(1)}_{\text{dt}} = \qty{-1815}{\V}$, $U^{(2)}_{\text{dt}} = \qty{-2275}{\V}$, $U^{(3)}_{\text{dt}} = \qty{-1805}{\V}$ and $U^{(4)}_{\text{dt}} = \qty{-2105}{\V}$ for (d-f).}
\label{fig:rotation}
\end{figure}

\subsection{Image Alignment}\label{subsec:ImageAlignment}
The previously discussed single-axis magnification can be used to implement an effective image rotation in the system. Taking advantage of the magnifying and focusing effect of the quadrupole field in the deflector region, an arbitrary atomic distribution can be stretched and aligned with the deflector's $x$- or $y$-axis. For demonstration, we employ a line of atoms rotated by $\alpha = \ang{45}$ to the deflector's $x$-axis with a length of \qty{100}{\um}, which shall be aligned to the $y$-axis on the imaging plane. 
The entire process is illustrated in Fig.~\ref{fig:rotation}a-c, which shows the evolution from the extraction region to the deflector region and finally to the detection plane. The initial distribution is shown in Fig.~\ref{fig:rotation}a, with the transverse electric field $E_{xy}$ illustrated by the arrows. Their length and color illustrate the absolute and y-directional field strength, respectively. The field stems from the radial field in Eq.~\eqref{eq:focusEffect}, which accompanies the electrodes' monopole terms near the optical axis. With the initial distribution being radially aligned, this field causes a (de)focusing along the optical axis. At the entrance of the deflector region the distribution is then still aligned at $\alpha = \ang{45}$, but radially stretched (cf. Fig.~\ref{fig:rotation}b). With a quadrupole field ($U_{\text{QP}}/U_{\text{dt}}=-0.1$) applied to the deflectors, the resulting force in the deflector region now causes the image to be strongly focused in $x$- and magnified in $y$-direction. This results in an effectively rotated image in the detection plane, now aligned with the y-axis (cf. Fig.~\ref{fig:rotation}c). Using this method, arbitrary initial distributions with $0< \left|\alpha\right| < \ang{90}$ can be aligned to either axis, $x$ or $y$. However, this method does not work at angles $\alpha = \ang{0}$ or $\ang{90}$. Here, we additionally need to employ a dipole term to the segmented extractor, as illustrated in Fig.~\ref{fig:rotation}d-f for an initial distribution at $\alpha = \ang{0}$. The dipole term is applied along the angle bisector in the $xy$-plane, effectively shifting the radial field from Eq.~\eqref{eq:focusEffect} diagonally outwards, such that the resulting field in y-direction (color of the arrows) shows a weak $x$-dependence (cf. Fig.~\ref{fig:rotation}d). It basically stems from the field inhomogeneity due to the finite electrode geometry and causes (besides a lateral shift) a small rotation of the initial distribution, which is now no longer aligned with the $x$-axis (cf. Fig.~\ref{fig:rotation}e). The quadrupole term of the deflector can then be used to magnify this small rotation, thereby aligning the cloud in the $y$-direction (cf. Fig.~\ref{fig:rotation}f). Conversely, the dipole term of the deflector can be used to shift the atoms back to the optical axis. 

Altogether, this demonstrates that any initial distribution can be aligned to either the $x$- or $y$-axis and motivates more complicated detector designs that fully harness the given possibilities.

\section{Image distortions}\label{sec:aberrations}
Similar to light optical systems, the performance of ion-optical imaging systems is generally limited by distortions. These include not only spherical and chromatic aberrations, but also astigmatism and the influence of magnetic fields in the system. 

\subsection{Spherical Aberration}
Spherical aberrations are inherent to rotationally symmetric imaging systems and can never be fully avoided \cite{scherzer1936ueber}. They describe how the distance from the optical axis affects the focal length. In order to minimize spherical aberrations, it is preferable for the ion trajectories to have small angles of incidence at the lens. Refraction primarily occurs in the \qty{1}{\mm} gap between the segmented and the conical electrode, where the potential landscape is changing rapidly. The applied voltages and the aperture size determine the focal length and aberration characteristics of the lens. In principle large apertures are preferable to minimize spherical aberrations \cite{szilagyi1986electrostatic}. However, as this reduces the focal length, large voltages at the electrode are required to achieve the desired focusing conditions. The voltages used are maximised considering the constraints mentioned in Sec.~\ref{subsec:constraints}.

\subsection{Chromatic aberration}
The main source for chromatic aberration is the starting energy distribution of the ions. Generally, the focal length for ions with lower energy decreases, while the refractive power of the lens is increased. This results in an image broadening for an object of given start energy distribution. For experiments with ultracold atoms, the starting energy is typically given by the clouds temperature and the momentum transfer during ionization. While the temperature effect amounts to $E_{\text{kin}} = k_B T = \qty{86.17}{\pico\eV\per\micro\K}$, the maximal energy spread $\Delta E$ due to the momentum transfer during ionization is negligible for field-ionization and amounts $\sim \qty{1}{n eV}$ for photo-ionization of $^{87}$Rb close to the ionization threshold \cite{stecker2017high}. Assuming cloud temperatures in the few \qty{}{\micro \kelvin} regime, the starting energy distribution is then typically below \qty{1}{\nano\eV}. This is negligible compared to the ion's kinetic energy acquired during detection, which ensures a 'sharp' image for almost all voltage configurations. 

\subsection{Depth of Field}
\begin{figure}[tbp]
	\includegraphics[width=0.47\textwidth]{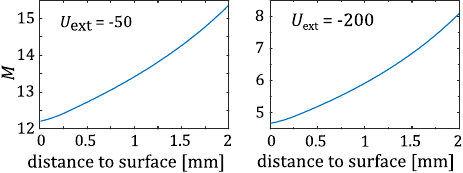}
	\caption{\textbf{Depth of field.} Magnification as function of the distance to the atom chip.  The extractor voltage was set to $U_{\text{ext}}=\qty{-50}{\V}$ and  $U_{\text{ext}}=\qty{-200}{\V}$, respectively.  For the simulation the remaining electrodes have been fixed to $U_{\text{chip}}=\qty{0}{\V}$, $U_{\text{con}}=\qty{-1000}{\V}$, $U_{\text{dt}}=\qty{-2000}{\V}$ and $U_{\text{CEM}}=\qty{-2300}{\V}$.}
	\label{fig:dof}
\end{figure}

The starting energy may also vary depending on the initial position of the ions. Although this effect is weak perpendicular to the ion-optical axis, it significantly contributes to aberrations in extended cloud formations along the optical axis, defining the objective's depth of field.
Atoms starting at greater distances from the surface travel through the ion optics with slightly lower velocity, resulting in increased magnification. Therefore, the magnification of the imaging system strongly depends on the ion's initial position relative to the surface. Figure~\ref{fig:dof} shows the magnification as function of the atoms starting position with respect to the surface for two different extractor voltages of \qtylist{-50;-200}{\V}, respectively. As expected, the magnification increases continuously with increasing initial distance to the surface. At a distance of \qty{2}{\milli\meter} this increase reaches an absolute value of $3$, as compared to the chip surface.

\subsection{Magnetic fields}
\begin{figure}[tbp]
	\includegraphics[width=0.47\textwidth]{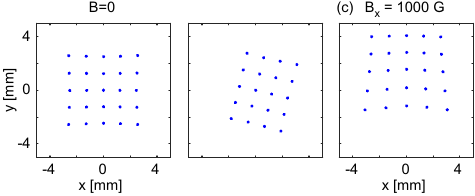}
	\caption{\textbf{Magnetic fields.} Imaging a $5\times 5$ ion grid with lattice constant of \qty{100}{\um} through the ion-optical system, for applied magnetic fields of (a) $B=\qty{0}{\G}$, (b) $B_{z}=\qty{3000}{\G}$ and (c) $B_{x}=\qty{1000}{\G}$. For the simulation the electrodes have been fixed to $U_{\text{chip}}=\qty{0}{\V}$, $U_{\text{ext}}=\qty{-50}{\V}$, $U_{\text{con}}=\qty{-1000}{\V}$, $U_{\text{dt}}=\qty{-2000}{\V}$ and $U_{\text{CEM}}=\qty{-2300}{\V}$.}
	\label{fig:bfeld}
\end{figure}

With the ion-optical system being designed for experiments with magnetic trapping of atoms on a chip, magnetic fields are unavoidable during operation. Apart from trapping, they also make for an appropriate tool to change the atoms s-wave scattering length via Feshbach resonances \cite{tiesinga1993threshold, inouye1998observation, courteille1998observation, chin2010feshbach}. Therefore, effects of ambient magnetic fields in the ion-optical imaging system are investigated. Figure~\ref{fig:bfeld} shows these effects, as calculated for a $5\times 5$ grid of ions separated by \qty{100}{\um}, starting \qty{100}{\um} below the chip surface. Without ambient magnetic fields (cf.~Fig.~\ref{fig:bfeld}a), the grid is imaged through the ion-optics at a central magnification of $M=12.25$, with only small  distortions of the pattern at the outer imaging regions. Figure~\ref{fig:bfeld}b shows the pattern for a magnetic field of \qty{3000}{\G} in direction parallel to the optical axis. As expected, this results in an image rotation around the optical axis, with a rotation angle of \qty{0.0045}{\degree\per\G}. Applying a constant field of \qty{1000}{\G} (cf.~Fig.~\ref{fig:bfeld}c) perpendicular to the ion-optical axis, causes the ion trajectories to be bent via the corresponding Lorentz force. This shifts the grid pattern in the imaging plane perpendicular to the magnetic field direction at a rate of \qty{1.5}{\um\per\G}. Due to the shift, the pattern is partially moved to the outer parts of the imaging region, where image distortions are increased. For typical magnetic field strengths at the atom chip of up to \qty{100}{\G} the magnetic field effects are thus negligible and do not alter the imaging quality.
\section{Calibration of detection efficiency}\label{sec:elion}
\begin{figure}
\centerline{\scalebox{1}{\includegraphics[width=0.48\textwidth]{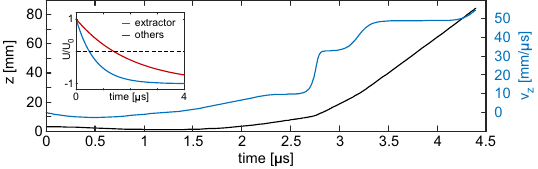}}}
\caption{\textbf{Detector calibration.} Coincidence measurement of an electron and ion from the same ionization event. At $t=0$ all voltages are set for electron detection ($U_{\text{chip}}=\qty{0}{V}$, $U_{\text{ext}}=\qty{+50}{V}$, $U_{\text{con}}=\qty{+1000}{V}$, $U_{\text{dt}}=\qty{+2000}{V}$, $U_{\text{CEM}}=\qty{+2300}{V}$). Starting from $t=0$, all electrodes are switched in sign on a time constant (1/e-time) of \qty{0.7}{\micro\second} for the extractor and \qty{2}{\micro\second} for the remaining electrodes. The trajectory and velocity of the ion, starting \qty{2}{mm} below the chip surface is shown. The electron is detected at $t=\qty{5}{\ns}$ and therefore unaffected from the voltage ramps, which are shown in the inset.}
\label{fig:ElektronIon}
\end{figure}

With future applications of the (Rydberg) correlation detector being strongly focused on the realization of Rydberg based multi-qubit quantum gates, measuring the gate fidelity is of specific importance. Therefore, the detection efficiency of the individual CEMs has to be calibrated carefully. This can be done by detecting both the electron and the ion from a single ionization process. As they are always produced in correlated pairs, the conditional probability for detecting an ion once an electron was detected, equals the ion detection efficiency \cite{stibor2007calibration}. The coincidence measurement can be achieved by first detecting the much faster electron and subsequently switching the CEM voltages, to detect the ion. For equal amplitudes of the electrode voltages, the trajectories of the negatively charged electron and the singly charged ion remain the same \cite{drummond1984ion}. However, the switching times have to be adapted to the electrons and ions flight times through the detector, which are typically very short. If the particles start at \qty{2}{mm} distance to the surface, the electron passes the entrance of the conical electrode after \qty{3}{\ns} and is detected after \qty{6}{\ns}. Meanwhile, the ions move only a negligible distance towards the chip surface, which they would reach only after \qty{570}{\ns}. After switching the voltages, the ion travels \qty{1.2}{\us} to the entrance of the conical electrode and \qty{2.4}{\us} until it is detected at the CEM. For real world application, however, the voltages have to be switched using voltage ramps.
In an appropriate switching scheme, using voltage ramps with exponential approach to the final value, the extractor has to be switched within the first \qty{0.7}{\us}, while the conical electrode, the deflection electrode and the front facet of the CEM-array are switched in \qty{2}{\us} (see inset of Fig.~\ref{fig:ElektronIon}). Although all voltage ramps start at $t=0$, the electron motion stays mainly unaffected, as it reaches the detector on the ns timescale. Following the switching scheme, Fig.~\ref{fig:ElektronIon} shows the ion position along the optical axis and the corresponding velocity over time. The ion reaches the detector \qty{4.4}{\us} after the electron has been detected. 

\section{Conclusion and Outlook}\label{sec:conclusion}
In conclusion, we have presented the design and characterization of a compact correlation detector for spatially separated ultracold Rydberg atoms. The system employs an electrode assembly in conjunction with an array of channel electron multipliers (CEMs) to enable the state-selective, spatially resolved detection of charged particles with high temporal resolution. Using individual CEMs ensures simultaneous detection at separate locations, making the system ideally suited for resolving correlations between Rydberg atoms. Simulations show that the magnification is primarily determined by the voltages applied to the extractor and conical electrodes, and values exceeding $12$ can be achieved with minimal aberrations. By segmenting the extractor and deflector electrodes, the detector can also achieve high single-axis magnifications up to $200$, as well as providing flexible control over trajectory rotation, the extraction region, and the placement of particles at the detector plane. The detector can be used to detect both ions and electrons, and the trajectories of the two are identical when the voltages are inverted. However, due to the lower kinetic energy (voltages) required to detect them, electrons provide an alternative detection mode with relaxed technical constraints. 

In the future, the system could be further improved by dynamically modulating the electrode potentials. For instance, a filtering technique could be employed that uses temporally varying electrode voltages to infer the initial velocities of particles, or to act as a velocity filter. Furthermore, switching between different CEMs dynamically could potentially reduce detector dead time and improve the temporal resolution.

Altogether, the presented detection system provides a versatile and compact platform for precision readout of neutral atom qubits, enabling correlated, state-selective detection of multiple Rydberg atoms. The integration in an atom chip experiment makes it well suited for advancing research in quantum information processing and cavity QED systems.

\section*{Acknowledgements}
The authors would like to thank C. Kalkuhl for helpful discussions on the manuscript. This work was supported by the Deutsche Forschungsgemeinschaft through FOR 5413, SPP 1929 GiRyd and QuantERA MOCA Proj. Nr. 491986552.
C.G. acknowledges support from the Evangelisches Studienwerk e.V. Villigst.


%

\end{document}